\documentclass[12pt,a4paper]{article}

\usepackage[T1]{fontenc}
\usepackage[utf8]{inputenc}
\usepackage[english]{babel}
\usepackage{lmodern}
\usepackage{microtype}
\usepackage[a4paper,margin=1in]{geometry}
\usepackage{setspace}
\usepackage{graphicx}
\usepackage{float}
\usepackage{bm}
\usepackage{mathtools}
\usepackage[numbers,sort&compress]{natbib}
\usepackage{caption}
\usepackage{subcaption}
\usepackage[colorlinks=true,linkcolor=blue,citecolor=blue,urlcolor=blue]{hyperref}
\usepackage{authblk}

\title{Physics-constrained machine learning for decoding multi-nanobubble configurations in graphene}

\author[1]{Jihye Kim}
\author[1]{Taegeun Song$^{*}$}
\author[2,3]{Nojoon Myoung$^{*}$}
\affil[1]{Department of Data Information Physics, Kongju National University, Gongju 32588, Republic of Korea}
\affil[2]{Department of Physics Education, Chosun University, Gwangju 61452, Republic of Korea}
\affil[3]{Well-Aging Medicare Institute and CSU G-LAMP Project Group, Chosun University, Gwangju 61452, Republic of Korea}

\date{}

\begin{document}

\maketitle

\begingroup
\renewcommand\thefootnote{}
\footnotetext{* Corresponding authors: tsong@kongju.ac.kr; nmyoung@chosun.ac.kr}
\endgroup

\begin{abstract}
Identifying multiple graphene nanobubbles from electronic spectra is challenging because their strain-induced features overlap. We develop a physics-constrained machine-learning framework that decodes nanobubble configurations from density-of-states (DOS) spectra. For spatially separated nanobubbles, previous full quantum-transport calculations established that the multi-bubble DOS is numerically equivalent to the normalized sum of the constituent single-bubble spectra. We encode this validated additive relation in a compact neural decomposition model. For each target spectrum, the basis coefficients are optimized independently, and the resulting weights directly identify the constituent geometries. The method accurately reconstructs configurations of increasing complexity and remains robust to repeated constituents, incomplete basis dictionaries, and simulated measurement noise. The framework provides an interpretable route for characterizing strain-engineered graphene nanostructures and may extend to other quantum materials with additive spectral responses.
\end{abstract}

\section*{Introduction}

Structural deformations in graphene, including uniaxial and biaxial strain\cite{mohiuddin2009uniaxial,si2016strain,naumis2017electronic,sahalianov2019straintronics}, ripples\cite{zwierzycki2014transport,dong2014theoretical,fan2017dominant,rakic2016large,deng2016wrinkled}, wrinkles and folds\cite{zhu2012structure,deng2016wrinkled,jun2025nanowrinkle}, and bubbles\cite{leconte2017graphene,lu2013properties,ghorbanfekr2017dependence,villarreal2021breakdown,park2023strain}, have been extensively studied because the associated synthetic gauge fields generate unusual electronic and transport properties. Recent experimental and theoretical studies have further examined the use of local strain for valley-dependent transport\cite{niu2012spin,hsu2020nanoscale,settnes2016graphene,zhai2018local}, electronic-band engineering\cite{qi2023recent,gui2008band,yang2022origami,wang2014strain}, and Dirac-fermion confinement\cite{kim2011interplay,neek2012nanoengineered,neek2012strain}. Consequently, the electronic responses induced by local structural deformations have opened new directions in condensed-matter physics. Among these localized structures, accurately identifying the geometries and spatial distributions of graphene nanobubbles is important for relating local strain fields to prospective device applications, including quantum dots, electron waveguides, and quantum interferometers.

Several experimental techniques have been used to characterize local strain structures, each offering different trade-offs between resolution and throughput. Optical methods such as dark-field imaging are suitable for rapidly screening nanobubble distributions over large areas\cite{georgiou2011graphene,stolyarova2007high}, while Raman spectroscopy enables non-invasive estimation of strain fields\cite{mohiuddin2009uniaxial,ni2008uniaxial}. Their spatial resolution is nevertheless limited by optical diffraction, which makes individual nanobubbles difficult to resolve. Atomic force microscopy (AFM) is a standard tool for topographic characterization\cite{klimov2012electromechanical}, but its spatial profiles are affected by tip convolution and it does not directly measure the local density of states (LDOS). Scanning tunneling microscopy (STM) and spectroscopy (STS) provide the spatial and spectral resolution required to investigate local electronic properties\cite{levy2010strain,lu2012transforming,xu2012atomic}; however, these methods have low throughput and require stringent environmental control and substantial data analysis. Other high-resolution techniques, including transmission electron microscopy (TEM)\cite{textor2018strategies} and scanning near-field optical microscopy (SNOM)\cite{fei2016ultraconfined}, are less suitable for routine characterization because of demanding sample preparation or instrumental complexity. An efficient framework for interpreting complex spectra and identifying multi-nanobubble configurations would therefore complement these experimental techniques.

Machine learning (ML) has emerged as a complementary approach for characterizing local strain effects. Data-driven methods have shown considerable promise in interpreting $dI/dV$ spectra\cite{oli2023atomic,lee2025rapid,zou2023deciphering,roccapriore2021revealing,narasimha2025uncovering}. For example, ML algorithms can predict the geometric parameters---height and width---of individual graphene nanobubbles from strain-correlated features in $dI/dV$ curves, which can approximate the DOS or LDOS under the appropriate measurement conditions\cite{song2021machine,nedell2022deep}. Although such approaches have been successful for isolated nanobubbles, conventional ML models face substantial challenges for configurations containing multiple nanobubbles\cite{kim2024neural}. Generic neural networks are often implemented as black-box models that rely on statistical pattern recognition without explicit physical constraints\cite{oviedo2022interpretable,dawid2020phase}. They may therefore have difficulty separating overlapping spectral contributions associated with distinct strain profiles. A physically constrained framework capable of decomposing these superposed spectra could reduce both the throughput bottleneck of high-resolution measurements and the burden of spectral interpretation.

Here, we develop a physics-constrained neural decomposition framework for identifying multi-nanobubble configurations from graphene DOS spectra. The method uses a dictionary of single-bubble spectra generated by quantum-transport calculations and embeds an additive relation that was previously validated by full multi-bubble simulations for sufficiently separated nanobubbles\cite{kim2024neural}. For each target spectrum, the basis coefficients are optimized independently rather than predicted by a global black-box regressor. We evaluate the framework for increasingly complex configurations, repeated constituents, incomplete basis dictionaries, and simulated measurement noise. The optimized weights provide a direct physical interpretation by identifying the constituent single-bubble spectra associated with the measured composite response.

\section*{Results}

\subsection*{Physics-constrained neural decomposition framework}

%%%%%%%%%%%%%%%%%%%%%%%%%%%%%%%%%%%%%%%%%%%%%%%%%%%%%%%%%% FIG1
\begin{figure}[htbp!]
    \centering
    \includegraphics[width=\linewidth]{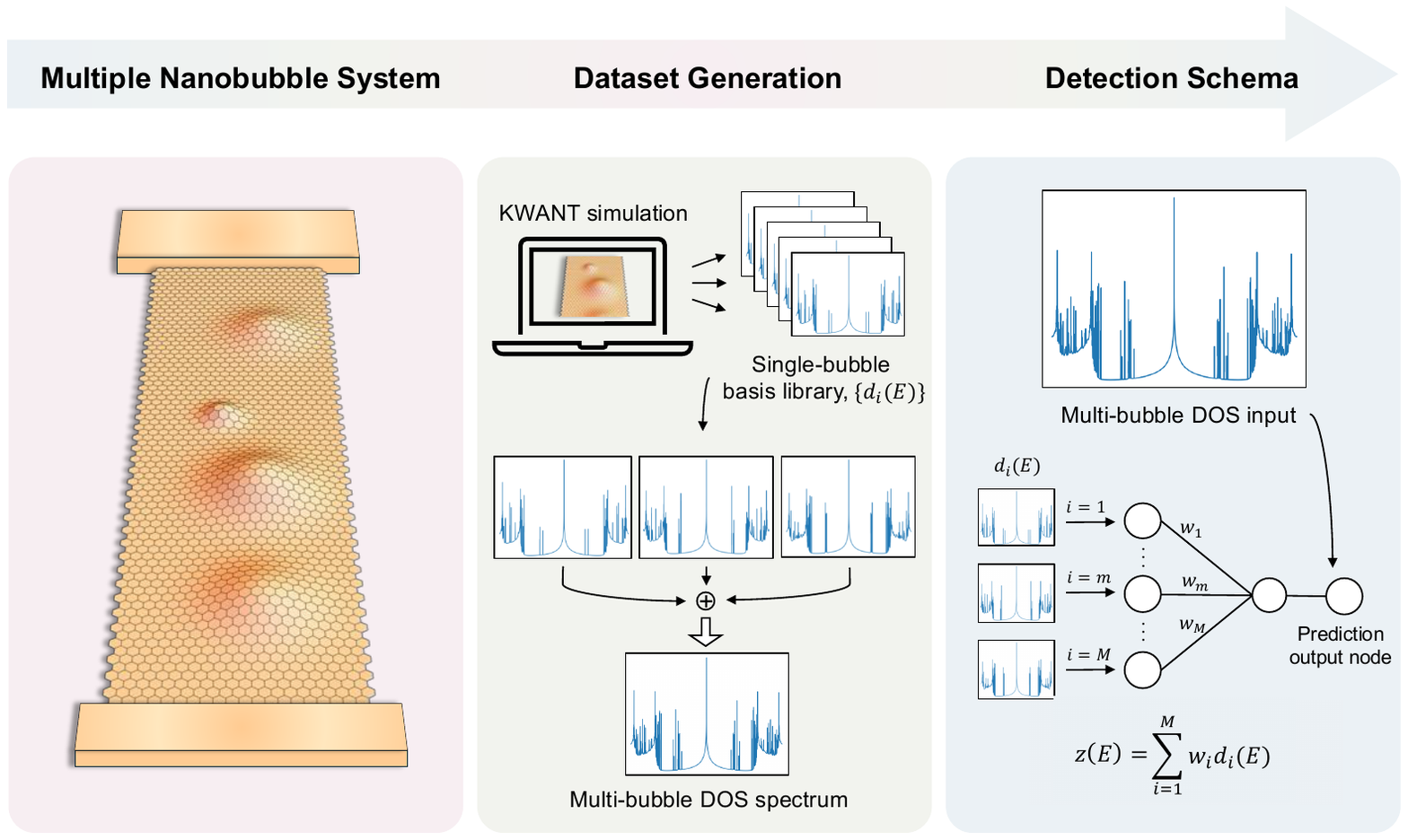}
    \caption{Schematic overview of the physics-constrained framework for identifying multi-nanobubble configurations in graphene. The left panel illustrates a graphene nanoribbon containing multiple nanobubbles, each characterized by its height $h$ and lateral width $\sigma$. The middle panel shows the dataset-generation procedure: single-bubble DOS basis spectra are obtained from quantum-transport simulations using the \textsc{Kwant} Python package and are combined as a normalized sum to construct composite multi-nanobubble spectra. This construction follows the previously verified numerical equivalence between full multi-bubble calculations and summed single-bubble spectra for sufficiently separated nanobubbles. The right panel presents the decomposition procedure, in which the basis coefficients are optimized for an individual composite DOS and the resulting weights reveal the corresponding nanobubble composition.}
    \label{fig1}
\end{figure}
%%%%%%%%%%%%%%%%%%%%%%%%%%%%%%%%%%%%%%%%%%%%%%%%%%%%%%%%%%

Figure~\ref{fig1} summarizes the proposed framework. Previous full quantum-transport calculations established that, when the nanobubbles are sufficiently separated, the multi-bubble DOS is numerically indistinguishable from the normalized sum of the corresponding single-bubble DOS spectra\cite{kim2024neural}. We use this validated additive relation as a physics-derived inductive bias rather than asking an unrestricted model to learn it from data. The inverse problem is therefore formulated as the decomposition of a composite spectrum into a physically generated basis dictionary.

The computational model is intentionally compact. The basis dictionary and aggregation rule are fixed, whereas a new coefficient vector is initialized and optimized independently for each target spectrum. A single aggregation layer forms a weighted combination of the candidate basis spectra, so that every optimized coefficient is associated with a physically defined nanobubble geometry. The procedure is therefore an optimization-based spectral decomposition implemented in a neural-network formalism, rather than a global pretrained regressor that predicts coefficients in a single forward pass. Restricting the hypothesis space in this manner reduces the opportunity for unphysical solutions, and the resulting weight distribution can be interpreted directly in terms of nanobubble composition.

The neural formalism serves as a differentiable implementation of this constrained inverse problem. Its effectiveness derives from the physically specified basis and the validated aggregation rule rather than from the universal approximation capacity of a deep network. Accordingly, the optimized parameters represent constituent weights rather than unconstrained latent features.

\subsection*{Reconstruction of composite DOS spectra}

%%%%%%%%%%%%%%%%%%%%%%%%%%%%%%%%%%%%%%%%%%%%%%%%%%%%%%%%%% FIG2
\begin{figure}[htbp!]
    \centering
    \includegraphics[width=0.8\linewidth]{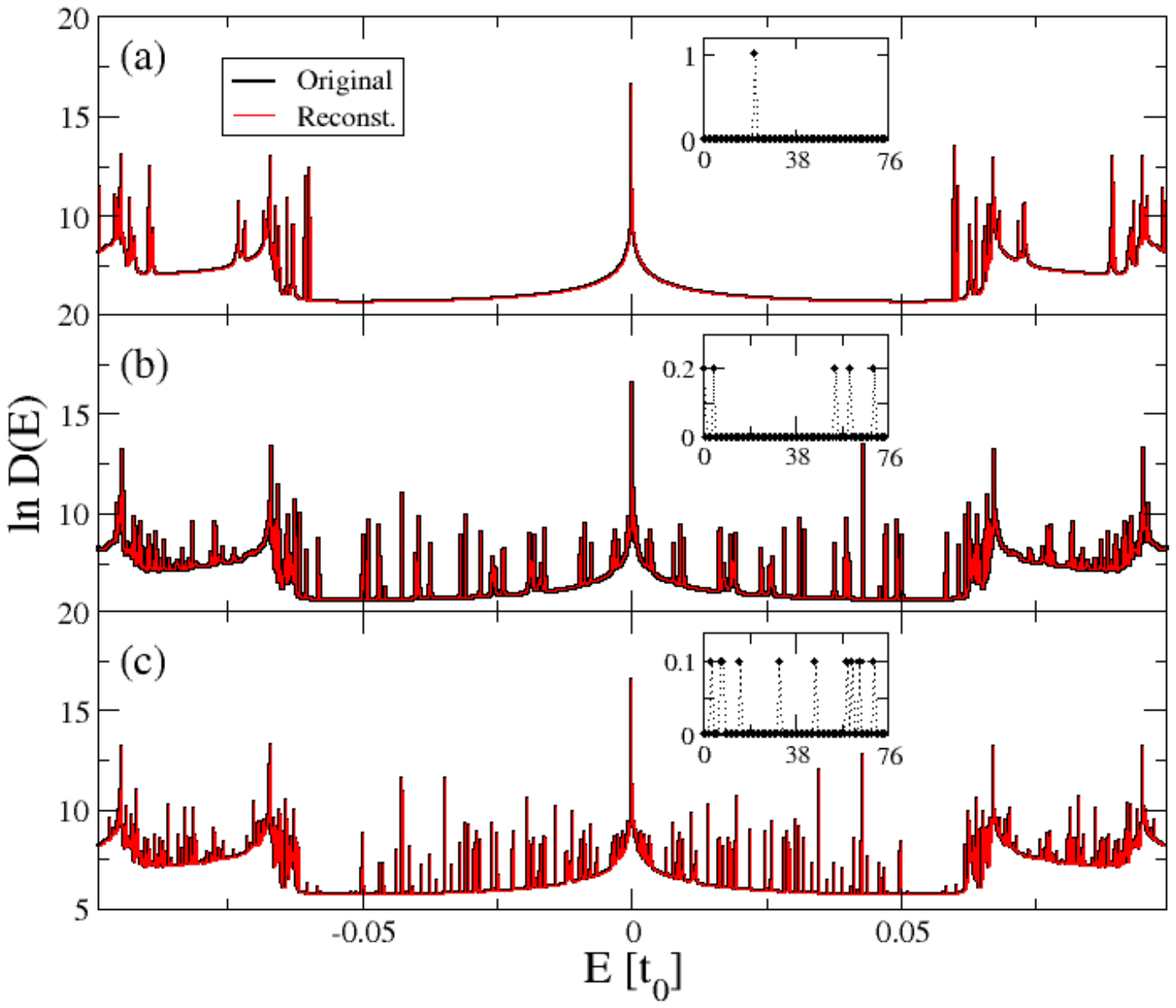}
    \caption{Reconstruction of composite DOS spectra for multi-nanobubble configurations. The original composite spectra (black) are compared with spectra reconstructed through spectrum-specific coefficient optimization (red) for systems containing (a) one, (b) five, and (c) ten nanobubbles. The insets show the corresponding optimized weight distributions over the single-bubble basis library. The close agreement between the curves indicates that the decomposition identifies and reconstructs the constituent contributions as the number of nanobubbles increases.}
    \label{fig2}
\end{figure}
%%%%%%%%%%%%%%%%%%%%%%%%%%%%%%%%%%%%%%%%%%%%%%%%%%%%%%%%%%

We first examine whether the proposed method can faithfully reconstruct composite DOS spectra of increasing complexity. Figure~\ref{fig2} compares the original composite DOS curves with spectra reconstructed by the model for representative systems containing one, five, and ten nanobubbles. In all cases, the reconstructed curves reproduce the principal resonance structures of the target spectra with high fidelity, including both isolated sharp peaks and crowded spectral regions. The agreement remains robust as the number of constituent nanobubbles increases, indicating that the additive basis representation retains sufficient capacity for the configurations considered here.

The insets of Fig.~\ref{fig2} provide direct evidence of the physical interpretability of the method. For the single-bubble case, the optimized weight distribution is strongly localized at the correct basis component. For the five- and ten-bubble cases, several nonzero weights emerge, reflecting the greater complexity of the underlying compositions. Importantly, the procedure does not merely reproduce the overall line shape; it identifies a sparse or structured set of physically meaningful basis spectra whose superposition yields the observed DOS\@. This behavior supports its interpretation as a constituent-decomposition method rather than a generic curve-fitting procedure.

\subsection*{Tests beyond idealized basis conditions}

%%%%%%%%%%%%%%%%%%%%%%%%%%%%%%%%%%%%%%%%%%%%%%%%%%%%%%%%%% FIG3
\begin{figure}[htbp!]
    \centering
    \includegraphics[width=\linewidth]{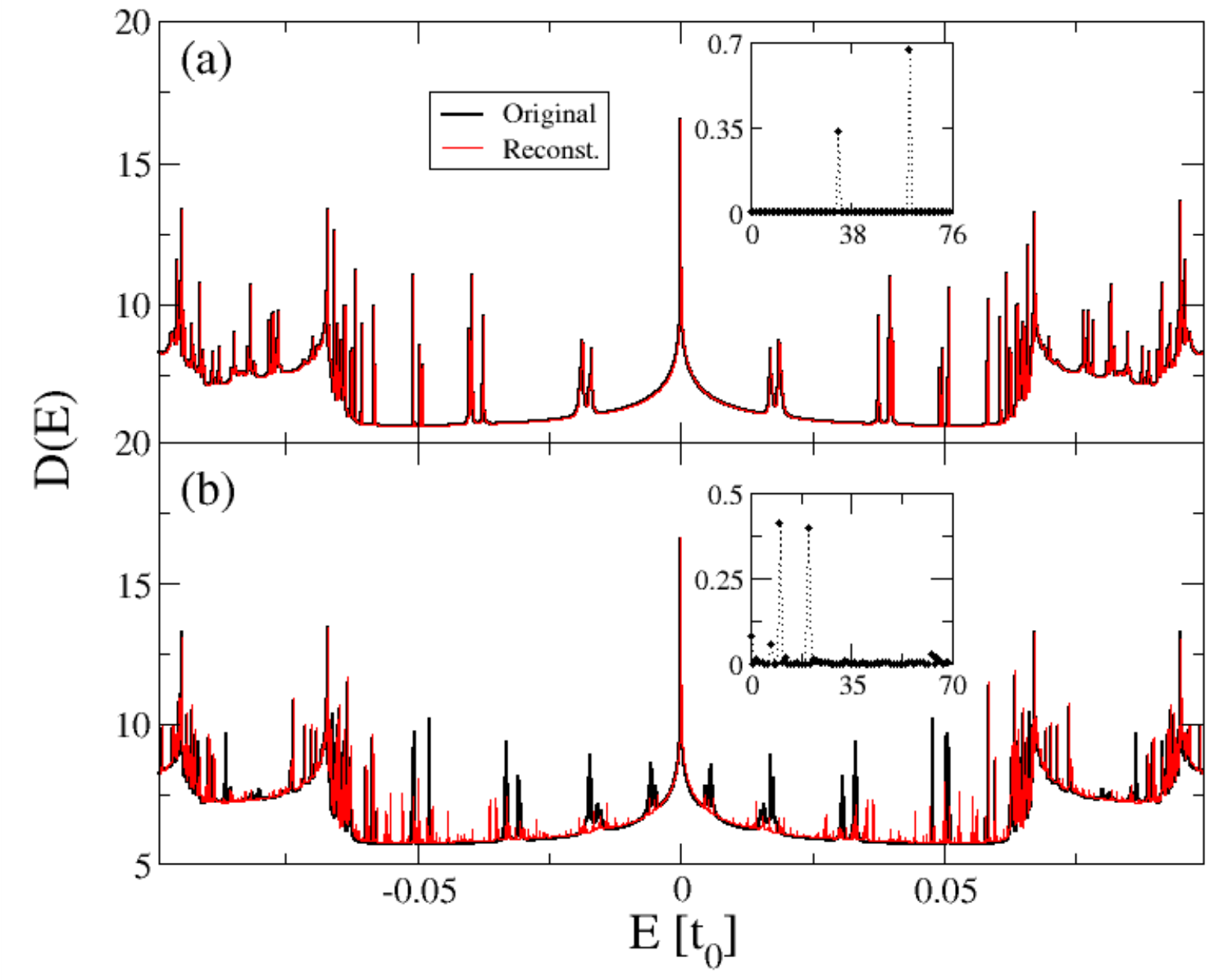}
    \caption{Tests beyond idealized basis conditions. (a) Reconstruction of an imbalanced configuration containing two identical nanobubbles and one distinct nanobubble, showing that the decomposition identifies repeated basis components through their fractional multiplicity weights. (b) Reconstruction with an incomplete basis dictionary, in which one of the three constituent nanobubbles is absent from the available dictionary; the number of basis channels is reduced from 76 to 70. In both panels, the original DOS is shown in black, the reconstructed DOS is shown in red, and the insets display the optimized basis weights.}
    \label{fig3}
\end{figure}
%%%%%%%%%%%%%%%%%%%%%%%%%%%%%%%%%%%%%%%%%%%%%%%%%%%%%%%%%%

To assess the method beyond ideal basis conditions, we consider two more stringent tests. In Fig.~\ref{fig3}a, an imbalanced configuration contains two identical nanobubbles and one distinct nanobubble. This case tests whether the model can detect both the presence of a basis component and its repeated contribution to the composite spectrum. The reconstructed DOS remains in close agreement with the original curve, and the optimized weights reproduce the multiplicity of the repeated constituent. The method can therefore distinguish compositionally imbalanced configurations without additional architectural complexity.

A more demanding situation is shown in Fig.~\ref{fig3}b, where one of the three constituent nanobubbles is excluded from the basis set. The full optimization array contains 76 basis channels, whereas the incomplete-dictionary test retains 70 channels. In this incomplete-dictionary scenario, the target spectrum cannot be represented exactly by the available library. Nevertheless, the model provides a reasonable reconstruction by selecting the closest admissible combination of known basis spectra. This result is relevant experimentally because a practical basis library is unlikely to be exhaustive. The physically constrained hypothesis space therefore provides a degree of tolerance to incompleteness in the prior dictionary.

\subsection*{Robustness against measurement-like perturbations}

%%%%%%%%%%%%%%%%%%%%%%%%%%%%%%%%%%%%%%%%%%%%%%%%%%%%%%%%%% FIG4
\begin{figure}[htbp]
    \centering
    \includegraphics[width=\linewidth]{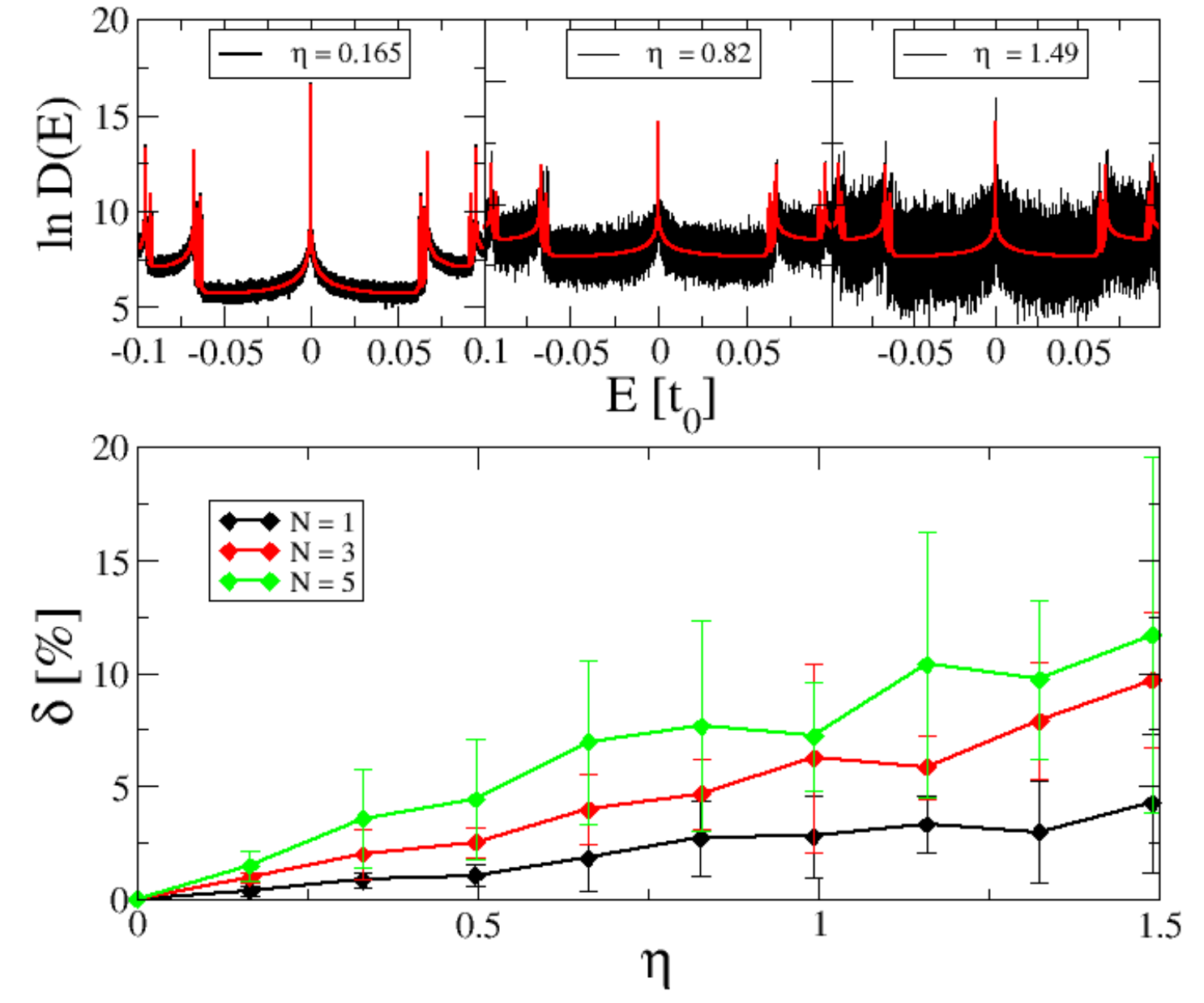}
    \caption{Robustness against additive Gaussian white noise in the log-transformed input spectra. Top: representative noisy log-DOS spectra obtained by adding Gaussian white noise after logarithmic transformation; the noisy inputs are shown in red. Bottom: mean normalized $L_1$ weight error $\delta_w$ as a function of the mean noise amplitude for configurations containing $N=1$, 3, and 5 nanobubbles. At each noise level, the averages are taken over 7, 6, and 8 independent realizations for $N=1$, 3, and 5, respectively; the vertical error bars denote the standard deviation of $\delta_w$. The error increases with both noise amplitude and configuration complexity, while the decomposition remains stable over the tested range.}
    \label{fig4}
\end{figure}
%%%%%%%%%%%%%%%%%%%%%%%%%%%%%%%%%%%%%%%%%%%%%%%%%%%%%%%%%%

We next investigate robustness to measurement-like perturbations by adding Gaussian white noise after the logarithmic transformation of the composite spectra. Figure~\ref{fig4} shows representative noisy log-DOS spectra and the corresponding normalized $L_1$ weight error $\delta_w$, defined in Eq.~\eqref{eq:weight_error}. The dominant resonances remain visible over a substantial noise range, whereas finer spectral features are gradually obscured as the noise amplitude increases. Even under these perturbations, the decomposition recovers the principal characteristics of the original spectra. The weight error increases gradually with noise amplitude and more rapidly for configurations containing larger numbers of nanobubbles, but the overall degradation remains moderate over the range considered.

The observed robustness follows from the structure of the model. Because reconstruction is restricted to the span of the physically motivated single-bubble basis spectra, the model is discouraged from fitting fluctuations that are inconsistent with the nanobubble dictionary. In this sense, the framework acts not only as a constituent predictor but also as a physics-guided spectral filter. The noise analysis supports the feasibility of applying the method to experimentally acquired DOS or $dI/dV$ data, for which instrumental noise and background fluctuations are unavoidable.

Taken together, the results show that the framework provides more than a numerical fit to a composite spectrum. Encoding the additive prior directly in the neural decomposition architecture yields a compact and interpretable model. The optimized weights identify the single-bubble basis components responsible for the composite DOS and thereby provide explicit physical information about the underlying multi-nanobubble configuration.

\section*{Discussion}

We have developed a physics-constrained machine-learning framework for identifying multi-nanobubble configurations in graphene from composite DOS spectra. The central feature of the method is the direct incorporation of a physically validated additive spectral relation into a minimal neural decomposition architecture. This design enables accurate reconstruction while retaining a transparent relation between the optimized weights and the constituent nanobubble geometries. The method remains effective for increasingly complex configurations, repeated components, incomplete basis dictionaries, and noisy spectra.

The scope of the method is set by the regime in which the additive relation applies. Prior full multi-nanobubble quantum-transport calculations showed that, for sufficiently separated nanobubbles, the normalized sum of the corresponding single-bubble DOS spectra agrees with the full multi-bubble result within numerical accuracy\cite{kim2024neural}. The present decomposition therefore uses a previously validated representation rather than an untested synthetic approximation. Strongly overlapping or interacting nanobubbles remain outside this regime and may require an extended dictionary containing interaction-dependent basis elements or a model with explicit nonlinear coupling terms. In addition, the present robustness tests use simulated Gaussian noise added in logarithmic spectral space, whereas experimental spectra may contain structured backgrounds, drift, tip-dependent matrix elements, and energy-dependent broadening. Experimental validation with tunneling-spectroscopy data will therefore be important for establishing the practical operating range of the framework.

Within these limitations, the approach offers a physically transparent route for converting complex spectra into constituent structural information. The same strategy may be useful for other two-dimensional and quantum materials in which measured spectra can be approximated by superpositions of localized structural or electronic motifs.

\section*{Methods}

\subsection*{Nanobubble geometry and DOS basis spectra}

The present study uses the same single-bubble DOS dataset reported in Ref.~\cite{kim2024neural}, generated from the quantum-Hall graphene model introduced in Ref.~\cite{song2021machine}. The graphene $\pi$-electron system is described by the nearest-neighbor tight-binding Hamiltonian
\begin{equation}
    H_{\mathrm{TB}}
    =-\sum_{\langle i,j\rangle} t_{ij}c_i^{\dagger}c_j+\mathrm{H.c.},
    \label{eq:tb_hamiltonian}
\end{equation}
where the unstrained nearest-neighbor hopping scale is $t_0\simeq3$~eV. The strain- and magnetic-field-dependent hopping elements $t_{ij}$, together with the device geometry, open leads, and boundary conditions, are implemented in the \textsc{Kwant} model exactly as in Refs.~\cite{song2021machine,kim2024neural}. The DOS spectra are calculated numerically with the \textsc{Kwant} implementation used in Refs.~\cite{song2021machine,kim2024neural}.

Each nanobubble is characterized by its maximum height $h$ and lateral width $\sigma$. The out-of-plane deformation centered at $\mathbf{r}_0$ is represented by
\begin{equation}
    z(\vec{r})=h\exp\!\left[-\frac{|\vec{r}-\vec{r}_0|^2}{2\sigma^2}\right].
    \label{eq:bubble_profile}
\end{equation}
The resulting $C_{3v}$-symmetric pseudomagnetic field is
\begin{equation}
    \vec{B}_{\mathrm{ps}}(\vec{r})
    =\nu\frac{\hbar\beta}{e a_0}
    \frac{h^2}{\sigma^6}\rho^3
    \exp\!\left(-\frac{\rho^2}{2\sigma^2}\right)
    \sin(3\theta)\,\hat{z},
    \label{eq:pseudomagnetic_field}
\end{equation}
where $\rho=|\vec{r}-\vec{r}_0|$, $\nu=\pm1$ labels the two valleys, $\beta=3.37$, and $a_0=0.146$~nm is the nearest-neighbor carbon--carbon distance. The graphene lattice constant is $a=\sqrt{3}a_0=0.246$~nm. The dataset contains $275$ single-bubble spectra sampled over $\sigma=20a,21a,\ldots,30a$ and $h=a,2a,\ldots,25a$. The raw DOS contains $20{,}001$ points over $-0.1\le E\le0.1$~eV and is interpolated to $10{,}001$ points over $-0.099\le E\le0.099$~eV. These spectra constitute the parent library from which the basis dictionary is selected.

\subsection*{Composite spectra and neural decomposition}

For a configuration containing $N$ spatially separated nanobubbles, let $n_i$ denote the multiplicity of the $i$th single-bubble basis spectrum $d_i(E)$, such that $N=\sum_{i=1}^{M}n_i$. We define the normalized composition weights as
\begin{equation}
    w_i^{\mathrm{true}}=\frac{n_i}{N},
    \qquad
    \sum_{i=1}^{M}w_i^{\mathrm{true}}=1.
    \label{eq:true_weights}
\end{equation}
In the separation regime established in Ref.~\cite{kim2024neural}, the full multi-bubble DOS is numerically equivalent, within the accuracy of the quantum-transport calculation, to the normalized superposition
\begin{equation}
    D_{\mathrm{comp}}(E)=\sum_{i=1}^{M}w_i^{\mathrm{true}}d_i(E).
    \label{eq:linear_composition}
\end{equation}
This convention represents repeated constituents through fractional multiplicity weights; for example, two identical nanobubbles and one distinct nanobubble have ground-truth weights $2/3$ and $1/3$, respectively. Spatially adjacent bubbles that coalesce into a single deformation are treated as one nanobubble and are not represented as independent additive constituents.

For a given target spectrum, the basis dictionary and the forward operation in Eq.~\eqref{eq:linear_composition} are fixed, whereas a new coefficient vector $\hat{w}$ is initialized and optimized independently. To compress the dynamic range of the DOS, both the target and reconstructed spectra are represented in logarithmic space. The model output is
\begin{equation}
    y_{\mathrm{pred}}(E)
    =\log\!\left[\epsilon+\sum_{i=1}^{M}\hat{w}_i d_i(E)\right],
    \label{eq:model_output}
\end{equation}
where $\epsilon$ is a small positive regularizer. The working optimization array used in the reported calculations has dimensions $50{,}000\times76$, with 76 basis channels. For the incomplete-dictionary test in Fig.~\ref{fig3}b, the corresponding array has dimensions $50{,}000\times70$.

Each target spectrum is treated as an independent coefficient-optimization problem. The optimization uses mini-batches of 128 samples, without separate validation or held-out test sets, because the objective is to represent the supplied target spectrum rather than to train a transferable coefficient-prediction network. The coefficients are obtained by minimizing the mean-squared error in logarithmic spectral space,
\begin{equation}
    \mathcal{L}_{\log}
    =\frac{1}{N_{\mathrm{s}}}
    \sum_{k=1}^{N_{\mathrm{s}}}
    \left[y_{\mathrm{pred},k}-y_{\mathrm{target},k}\right]^2,
    \label{eq:log_mse}
\end{equation}
where $N_{\mathrm{s}}$ is the number of samples in the optimization array. The loss is minimized using the Adam optimizer with an initial learning rate of $10^{-3}$. Optimization is run for at most $10{,}000$ epochs. Early stopping is applied when the optimization loss does not improve for 100 epochs, and the \texttt{ReduceLROnPlateau} scheduler lowers the learning rate by a factor of 0.5 after 50 epochs without improvement. The coefficients are initialized using the Xavier (Glorot) uniform scheme and clipped to the interval $[0,1]$ throughout optimization. No fixed random seed is imposed. Thus, the procedure is a spectrum-specific optimization-based decomposition implemented in a neural-network formalism, rather than a forward evaluation of one globally pretrained coefficient-prediction network. Because every coefficient is tied to a specified basis spectrum, the optimized weight vector directly identifies the constituent nanobubble geometries.

\subsection*{Noise model and evaluation}

To emulate measurement uncertainty, Gaussian white noise is added after the logarithmic transformation of the clean composite spectrum. Defining
\begin{equation}
    y_{\mathrm{clean}}(E)=\log\!\left[\epsilon+D_{\mathrm{comp}}(E)\right],
\end{equation}
the noisy target is
\begin{equation}
    y_{\eta}(E)=y_{\mathrm{clean}}(E)+\eta f(E),
    \label{eq:noise}
\end{equation}
where $f(E)$ is sampled independently from the standard normal distribution $\mathcal{N}(0,1)$ and $\eta$ controls the noise amplitude. No additional logarithmic transformation is applied after the noise is added.

For each plotted noise level in Fig.~\ref{fig4}, 7, 6, and 8 independent realizations are used for configurations containing $N=1$, 3, and 5 nanobubbles, respectively. The horizontal coordinate is the mean noise amplitude, and the vertical coordinate is the mean percentage error. The vertical error bars denote the standard deviation of the percentage error over the corresponding realizations.

The composition error is quantified by the normalized $L_1$ deviation of the optimized weights,
\begin{equation}
    \delta_w
    =100\,
    \frac{\sum_{i=1}^{M}|\hat{w}_i-w_i^{\mathrm{true}}|}
         {\sum_{i=1}^{M}|w_i^{\mathrm{true}}|}.
    \label{eq:weight_error}
\end{equation}
For the normalized ground-truth weights in Eq.~\eqref{eq:true_weights}, the denominator equals unity. This definition remains finite for basis elements that are absent from the target configuration and accommodates repeated constituents through their fractional multiplicity weights.

\section*{Data availability}
The simulation datasets generated and analyzed during the current study are available from the corresponding authors upon reasonable request.

\section*{Code availability}
The computational codes used for quantum-transport simulations, dataset generation, basis construction, and spectrum-specific coefficient optimization are available from the corresponding authors upon reasonable request.

\section*{Author contributions}
N.M. conceived the study, developed the physical model, performed the quantum-transport simulations, generated the electronic spectra, and interpreted the results. T.S. developed the machine-learning methodology, designed the physics-constrained framework, and supervised the computational analysis. J.K. constructed the datasets, implemented the decomposition framework, performed the spectrum-specific coefficient optimization and robustness analyses, and analyzed the data. All authors discussed the results, contributed to the scientific interpretation, and reviewed the manuscript.

\section*{Competing interests}
The authors declare no competing interests.

\section*{Acknowledgements}
This work was supported by National Research Foundation of Korea (NRF) grants funded by the Ministry of Science and ICT (MSIT) under Grant Nos. RS-2025-00557045 and RS-2026-25497242. This work was also supported by the G-LAMP Program through the NRF, funded by the Ministry of Education (MOE), under Grant No. RS-2023-00285353. The authors further acknowledge support from the Quantum Flagship Program funded by MSIT through the Institute of Information and Communications Technology Planning and Evaluation (IITP) under Grant No. RS-2025-25464832, Chosun University (2025), and a 2024 research grant from the Kongju National University Industry--University Cooperation Foundation.

\bibliographystyle{naturemag}
\bibliography{PMLMultiNB}

\end{document}